# Gene cluster analysis method reliably identifies horizontally transferred genes and reveals their involvement in operon formation

**Keiichi Homma[*], Satoshi Fukuchi[*], Yoji Nakamura[†], Takashi Gojobori[*,‡], and Ken Nishikawa[*]**

[*]Center for Information Biology-DNA Data Bank of Japan, National Institute of Genetics, Research Organization of Information and Systems, Mishima, Shizuoka, 411-8540 Japan; [†]Graduate School of Information and Technology, Hokkaido University, Kita 14, Nishi 9, Kita-ku, Sapporo, 060-0814 Japan; [‡]Japan Biological Information Research Center, National Institute of Advanced Industrial Science and Technology, AIST Tokyo Waterfront, 2-42 Aomi, Koto-ku, Tokyo 135-0064, Japan

Abbreviations: HT, horizontally transferred; HGT, horizontal gene transfer; ESS, *Eschericia*, *Shigella*, and *Salmonella* species used in this study; GHR, gene-homologous region; ROC, reduced ortholog cluster; LCA, last common ancestor

The formation mechanism of operons remains controversial despite the proposal of many models. Although acquisition of genes from other species, horizontal gene transfer, is considered to occur, definitive concrete cases have been unavailable. It is desirable to select horizontally transferred genes reliably and examine their relationship to operons. We here developed a method to identify candidates of horizontally transferred genes based on minimization of gene cluster insertions/deletions. To select a benchmark set of positively horizontally transferred genes against which the candidate set can be appraised, we devised another procedure using intergenetic alignments. Comparison with the benchmark set of horizontally transferred genes demonstrated the absence of a significant number of false positives in the candidates, showing that the method identifies horizontally transferred genes with a high degree of confidence. Horizontally transferred genes constitute at least 5.5% of the genes in *Escherichia*, *Shigella*, and *Salmonella* and ~46% of which originate from other γ−proteobacteria. Not only informational genes, but also operational genes (those involved in housekeeping) are horizontally transferred less frequently than expected. A gene-cluster analysis of *Escherichia coli* K-12 operons revealed that horizontal transfer produced four entire operons and expanded two operons, but deletion of intervening genes accounts for the formation of no operons. We propose that operons generally form by horizontal gene transfer. We further suggest that genes with related essential functions tend to reside in conserved operons, while genes in nonconserved operons generally confer slight advantage to the organisms and frequently undergo horizontal transfer and decay.

## Introduction

With three major models having been proposed (1), the evolutionary origin of operons in bacterial genomes still remains controversial. The Fisher model (2) states that operons form because the proximity of the co-adapted genes reduces the probability of obtaining unfavorable combinations of genes by recombination. Although this model explains operons whose constituent genes encode physically interacting proteins, the observed frequency of recombination is not high enough to justify the co-adaptation of genes (3). The co-regulation model (4) instead hypothesizes that operons facilitate coordinated expression and regulation. Though supported by the tendency of functionally related genes to reside in the same operons (5), this model fails to explain the existence of many operons containing genes of unrelated functions (6). Finally the selfish operon model (6) postulates that operons allow propagation of functionally related genes via horizontal transfer. The model proposes that horizontal transfer of a gene cluster containing weakly selected genes followed by deletion of intervening genes leads to operon formation. Despite the attractiveness of the gradual operon formation mechanism, the predictions that essential genes are not concentrated in operons and that operons frequently undergo horizontal transfer were not borne out by recent analyses (7, 8). We test the last two models through examinations of real operon formation events.

   All the hitherto available methods to identify horizontally transferred (HT) genes contain possible errors, making it impossible to draw definitive conclusions on the importance of horizontal gene transfer (HGT) and its characteristics. Besides the probabilistic nature, methods based on nucleotide compositions are ineffective in selecting short HT genes, those from phylogenetically close species whose compositions do not appreciably differ, and those whose compositions became indistinguishable from that of a host through amelioration (9, 10). For the same reasons compositional approach often leads to erroneous identification of HT genes (11-13). Phylogenetic methods share the first two weaknesses with compositional methods and often produce false identification because the smallness in the number of genes one examines in HT gene selection often produces incorrect phylogenetic trees (14), especially when the genes of close species are involved. The third type of methods determines HT genes from the presence/absence of genes and is based on the idea that two or more independent deletion events are less likely to occur than one insertion event and identify HT genes by the presence/absence of homologs in bacterial phyla (8, 15-17). They may miss HT genes if there are paralogues and fail to assess the extent of false positives. Moreover, the assessment of insertions/deletions (indels) on individual gene basis often fails to correspond to real-world indels of gene clusters. We here developed a novel method using the assessment of gene cluster alterations to identify HT genes with high reliability and suggest a new operon formation mechanism.

## Results and Discussion

**Clusters of orthologous genes.** It is more natural to analyse genome alterations in terms of gene clusters than individual genes, as genes usually undergo insertion and deletion in clusters. For this purpose, we first determined orthologs by identifying syntenic regions in eight closely related γ-proteobacterial species of the Enterobacteriaceae family (designated as the ESS species) whose genomes have been wholly sequenced: *Escherichia coli* K-12 (designated as ecol0), O157:H7 Sakai (ecol1), CFT073 (ecol3); *Shigella flexneri* 2a 301(sfle0), 2a 2457T (sfle1); *Salmonella typhi* CT18 (styp0), *Salmonella enterica* Typhi Ty2 (styp2), and *Salmonella typhimurium* LT2 (styp1). Though most genes were identified as orthologs, some genes show incomplete synteny or none at all (Fig. 1*a*). To take regions unannotated as genes into consideration, we identified intergenetic regions with homology to genes and designated them as gene-homologous regions (GHRs) and present examples in the figure. We regard a continuous stretch of genes with an identical presence/absence pattern among the ESS species as a reduced ortholog cluster (abbreviated as ROC, genes of the same color in the figure).

**HT gene identification method.** We developed the <u>Min</u>imum <u>Gene</u> <u>C</u>luster <u>Ind</u>el (MinGeneCIDE) method to identify HT genes through analyses of gene cluster alterations based on the following idea; if all the possible pathways to account for the current states in the ESS species in the minimum number of indels involve an insertion of a ROC, the insertion of the cluster after the last common ancestor of the ESS species (LCA) is probable. After expressing the presence and absence of each ROC by 1 and 0, respectively (Fig. 1*b*), we followed the MinGeneCIDE method (see Materials and Methods). We found 798 ROCs containing 2,451 groups of orthologous genes that exist in all the ESS species (colored orange and green in Fig. 1*b*) and call them conserved ROCs. We assume conserved ROCs not to have undergone indels after the LCA, in accordance with the minimum indel principle. As we hence need to analyse only those bits sandwiched by conserved ROCs (blue and pink-tinted cells), we represent the state of each species in a genome region with the exclusive use of such bits (e.g., 01 for ecol1 in the given example). There are 3,313 nonconserved ROC s in all.

The phylogenetic tree (Fig. 1*c*, designated as tree S,) is that inferred from genome conservation (18) and genome rearrangement (inversion) distances (19) and is identical to the tree based on the 16S rRNA sequences. Methods based on ortholog sequences (20) generally generate the alternative tree with slfe0 and sfle1 nesting with ecol0 (termed tree A, Fig. 2). As combined distances between genome rearrangements and ortholog sequences infer tree S (details of which will be published elsewhere), however, we regard tree S to represent the true phylogeny (see further discussion below). All the possible state(s) at each node are determined from the farthest branches of the phylogenetic tree (Fig. 1*c*). In this example, two minimum indel pathways exist with two indels each (starred states in one pathway, daggered states in the other, see Materials and Methods). The second ROC represented by the second digit (colored pink in Fig. 1*a*, *b*) indicates insertion in both cases (arrows at bottom left, Fig. 1*c*) is thus considered inserted. By contrast the first ROC is not classified as inserted because it is considered ambiguous; it may be either deleted (red arrow at top right) or inserted (green arrow at top middle) depending on the pathways. We then thought about a way to remove inserted genes caused by translocation or duplication to get HT genes; although inserted genes as a result of translocation or duplication generally have easily identifiable close homologs within the ESS species, some of the homologs may have been lost or undergone rapid evolution. We therefore attribute an inserted ROC to translocation or duplication if the best BLASTP hits in the GTOP database (21) of a majority of genes in the ROC are in the ESS species, and otherwise consider it to be horizontally transferred from other species. From the group of inserted genes, we exclude over-annotated genes. (A gene with at least one GHR but no other genes in the orthologous cluster in the ESS species is regarded as an over-annotated gene if it has no homologs in the GTOP database. There are 413 such genes.) The inserted ROCs were subjected to this Majority Rule screening to yield 2,016 HT gene candidates (5.5% of total, Table 1), including the pink-colored genes in Fig. 1. We note that the species breakdown of HT gene candidates may not faithfully reflect the real distribution, as the identifiable fraction by the MinGeneCIDE method is dependent on the tree topology.

**Selection of positively HT genes.** To test the reliability of the MinGeneCIDE method, we selected a benchmark set of positively HT genes by the Intergenetic Region Alignment (INTEGRAL) method, an automated method based on intergenetic alignments. From gene clusters, we first identify those that could have been inserted (e.g., the red genes in Fig. 3*a*) and examine if we can verify the insertion. Although simple comparison of two genomes yields gene clusters that are present in only one of them, one insertion in the species and one deletion in the other species explain the difference equally well. We therefore checked if we could verify insertions based on the conjecture that the boundaries of gene clusters must be different in general if two independent deletion events had occurred. If the intergenetic regions of spp. 2 and 3 can be aligned and nucleotides N2 and N3 are exactly matched (Fig. 3*b*), then we consider that the occurrence of two independent deletions in the two lineages is highly unlikely and X -Y is probably an inserted gene cluster. (It is theoretically possible to explain the present state by postulating the presence of W-X-Y-Z orthologs at LCA; introduction of W-Z homologous segments to spp. 2 and 3 and subsequent homologous recombinations may delete the X-Y segment at the identical location in the two species (22). However, we consider the occurrence of such cases extremely rare; nearly identical sequences must be horizontally transferred twice independently, once between node B and sp. 2 and another between node A and sp. 3, and the displaced homologous segment must be fixed in both species.) Although most genuine insertions are presumably not verifiable by this method due to the generally high mutability of intergenic regions, we could verify the insertions in 99 gene clusters consisting of 251 genes. They were then subjected to the Majority Rule screening to yield 24 gene clusters containing 116 positively HT genes (Table 2), including the red ROC in Fig. 3*a*.

**Characteristics of HT gene candidates.** To estimate how many false positives the set of HT gene candidates contains, we investigated the probable origins. The species in which the best BLAST hit of a HT gene candidate belongs to is regarded as the most likely origin. A gene that existed at the LCA of the ESS species is very likely to have orthologs in other γ−proteobacteria. In fact, 91.7% of the genes in conserved ROCs, which we consider to be native genes, have best hits in γ−proteobacteria (Table 3). Therefore, if the set of HT gene candidates contains some false positives, the apparent proportion of those originating from γ−proteobacteria must be higher than that of the set of positively HT genes. The fact the observed fraction of HT gene candidates whose probable origins are γ−proteobacteria (45.0%) is nearly identical to that of positively HT genes (45.7%; Table 2) therefore shows that the set does not contain a significant number of non-HT genes. It is theoretically possible that the HT identification method selects HT genes from phylogenetically close species less effectively and therefore gives a lower fraction of HT genes from proteobacteria, but incorporates some false positives to exactly compensate for the difference. However, the general similarity of the fraction distributions in Table 2 (no statistical difference at $P < 0.01$) makes this interpretation unlikely.

**Analyses assuming the alternative phylogenetic tree.** We selected HT gene candidates postulating tree A (Fig. 2) with the maximum number of combinations of states at nodes A, B, and C set as 50,000, so that computation can be completed in reasonable time. Computation assuming tree S with the same restriction on computation reduces the number of HT candidates from 2,016 to 1,973, but does not significantly affect the distribution of probable origins (data not shown). Thus the number of HT candidates assuming tree A, 4,667, represents a 2.37-fold increase from the corresponding figure supposing tree S. 102 positively HT genes were also chosen by the INTEGRAL method hypothesizing tree A. The distributions of probable origins (Table 3) show that there is a significant increase in the fraction of HT gene candidates from γ−proteobacteria as compared with that of the positively HT genes. If we assume that the conserved ROCs contain no HT genes, the fraction of such genes with best BLAST hits to γ−proteobacteria, 91.7%, corresponds to that of 100% false positives. With the further assumption that there are no false positives in the positively HT genes, 10.5% of the HT candidates are estimated to be false positives by interpolation. One problem in this error estimate is the dependence of positively HT genes on phylogeny. If we instead use the value of the positively HT genes identified assuming tree S (Table 2), the false positive rate increases to 13.3%. In either case, more than 10% of the HT gene candidates assuming tree A are false positives. We interpret both the drastic increase in the number of HT candidates and the inclusion of false positives as reflection of the wrong phylogeny; to account for the presence and absence of orthologs along the erroneous phylogenetic tree, many native genes were mistakenly classified as HT genes.

**Analyses of HT genes.**    Another look at the distribution of the probable origins of HT genes (Table 2) reveals that HT genes preferentially originate from phylogenetically close species: more than 45% from γ−proteobacteria other than the ESS species. The skew in the origin distribution of the identified HT genes to phylogenetically close species is reasonable, as genes of similar species are more likely to function in host cells due to similarity in transcription factors, codon usage, protein repertoire, and other factors, and therefore have higher fixation probability. Although all the genes in the two *pap* pilus operon-containing pathogenicity islands present in ecol3, but not in ecol1 (23) were selected as inserted genes, only a few of them were classified as HT gene candidates. This is because the Majority Rule conservatively rejects those that could have been translocated or duplicated. None of the HT genes in ecol0 in our list is classified as essential genes, while proportionally 10 are expected to be.

The HT genes were classified into the main functional roles (24) (Fig. 4). Intriguingly, genes involved in viral functions and mobile and extrachromosomal element functions are horizontally transferred significantly more frequently than expected ($P<0.01$), supporting the idea that HGT is frequently mediated by bacteriophages (25). The complexity hypothesis (26) states that informational genes are horizontally transferred at lower frequency than others, while those involved in housekeeping (operational genes) undergo HGT more frequently. Our data confirm that informational genes (in blue letters in Fig. 4) are rarely horizontally transferred. However, most of the categories corresponding to operational genes (in orange characters) contain HT genes at lower frequency. Overall, both informational and operational genes are horizontally transferred more rarely than expected ($P<0.01$).

**Characteristics of the MinGeneCIDE method.** Although the MinGeneCIDE method may miss some HT genes, it identifies HT genes with high reliability. We thus state that at least 5.5% of genes in ESS species are HT genes. By contrast simple genome comparison often leads to overidentification of HT genes. Approximately 47% of the HT genes we identified were also identified as HT genes by a method based on nucleotide compositions (10) (the last column in Table 2). The coverage, however, depends on the origins; the shorter the phylogenetic distance between the ESS species and the species of origin, the less likely the HGT is identified by the nucleotide composition method. This result makes sense as the MinGeneCIDE method is independent of the species of origin, while methods based on nucleotide compositions are prone to under-identify HT genes from phylogenetically close species (10).

**Conservation of operons.** We then examined operon conservation. Out of a total of 256 experimentally verified operons in ecol0, 164 were found to be conserved in all, while 92 operons are not conserved in at least one of the ESS species (Table 4). There are a handful of cases in which operons may be preserved in modified forms in both species despite the non-conservation of some constituent genes (Fig. 5). In the example, the orthologs of the ecol0 *rhaD* gene in styp0 and styp2 have deletions in the middle. Nevertheless, the conservation of the two upstream orthologs in the two species leaves the possibility that the remaining operon is still functional.

However, cases in which operons not conserved in species X may be functional in both ecol0 and X are few in number. Thus an operon creation in ecol0 or an operon loss in species X must account for most cases of non-conserved operons. Since we have no reason to expect a particular prevalence of operons in ecol0 as compared with other ESS species and the LCA, it is likely that approximately half, or ~20, of the operons nonconserved in *Salmonella* (Table 4) are attributable to operon creation between the LCA and ecol0, while the rest are cases of operon loss between the LCA to *Salmonella*. (Although ~20 cases of operon loss from the LCA to ecol0 and ~20 cases of operon creation from the LCA to *Salmonella* are also expected, these are not detectable by analyses based on operons currently present in ecol0.) Curiously, 17.2% of genes in the conserved operons are essential, while essential genes comprise only 7.4% in the nonconserved operons, nearly the same as the average frequency, 8.2%. Thus, while essential genes are indeed preferentially found in operons overall (5, 7), the nonconserved operons do not contain essential genes more frequently than the average.

Can we identify concrete cases of operon creation? For this purpose, we first examined the nonconserved operons for evidence of intervening gene deletion that produced new operons after the LCA (6) (Fig. 6*a*). Such genes in ROCs must generally have orthologous genes in separate sections in ESS species other than ecol0, because of the presence of genes orthologous to the deleted genes. We searched for such nonconserved operons and found only three cases: the *rst*, *purF*, and *aga* operons. In the *rst* operon (Fig. 7*a*), the MinGeneCIDE method identified an insertion event between nodes B and E resulting in the purple ROC. The existence or absence of the red ROC in the LCA is ambiguous; it may have been present in the LCA and subsequently deleted between nodes A and B, or absent in the LCA and inserted between nodes A and C. The latter is much more probable as the orthologs of *rstA* and *rstB* are contiguous in a number of species, including a very close ESS relative, *Yersinia pestis*, while there are no species in GTOP containing the homologs of the red genes in between. Therefore the *rst* operon was in all likelihood not formed by the deletion of the purple or red ROC. Similar observations also make it unlikely that the *purF* operon was formed by deletion. Lastly the *aga* operon contains an extra red ROC (Fig. 7*b*) in four species. The forward direction, Swiss-Prot annotation, and short intergenetic distances of the red genes all support the notion that they are part of the operon in these species. Furthermore the report that *agaW* and *agaA* in ecol0 are pseudogenes with the C-terminal and N-terminal half deleted, respectively (27), supports the idea that the red ROC was deleted together with fragments of the upstream and downstream genes in ecol0 (red dotted line in Fig. 7*b*). Thus this is probably a case of an operon that existed at node B losing some of the constituent genes, rather than operon formation by shedding intervening genes. We therefore are left with no operons that were plausibly formed by deletion of intervening genes in the past ~100 million years (28) from the divergence of *Salmonella* and *Escherichia*.

On the other hand the MinGeneCIDE method identified 24 HT genes of ecol0 in four operons: the entire *glc*, *fecABCDE*, *cynTSX*, and *lac* operons, *wbbK*, *wbbJ*, *wbbI*, *rfc*, *glf*, and *rfbX* in the *rfb* operon, and *chpR* and *chpA* in the *relA* operon (Fig. 3*a*). We note that the *lac* operon was previously suggested to be horizontally transferred (29). Though the number of cases is less than the expected number assuming proportionality, 40, their existence is undeniable, especially because the last two are positively HT genes. The horizontal transfer of operons as a whole (Fig. 6*b*) was proposed previously (22). The *relA* operon is probably a mixture of HT and non-HT genes because the *relA* gene is in a conserved ROC (Fig. 3*a*) and therefore in all probability existed in the LCA. In the *relA* operon, the *relA* gene located upstream in the operon encodes ATP:GTP 3'pyrophosphotransferase required in the stringent response to amino acid deprivation, while the downstream two genes, *chpR* and *chpA*, constitute an antitoxin/toxin system and were proposed to be responsible for programmed cell death (30). Both the disparate role the antitoxin/toxin system plays and the existence of an additional promoter between *relA* and *chpR* (31) support the notion that the *chpR* and *chpA* are HT genes. There are thus concrete cases of operon expansion by HGT (Fig. 6*c*).

The MinGeneCIDE method for HT gene identification differs from existing methods in that it analyses alterations of gene clusters instead of those of individual genes. This approach makes it possible to incorporate GHRs, which are important vestiges of orthologs, in its analysis. According to this method, insertions or deletions of adjacent gene clusters are counted as one indel, as they should be. The two salient features cannot be incorporated in HT gene identification methods based on analyses of individual genes. The gene cluster approach also makes it possible to positively identify HT genes by insertion-verification and thereby to test the reliability of selected HT genes. Only four cases of operon creation (the *glc* and *fecABCDE*, *cynTSX*, and *lac* operons) from the LCA to the present was uncovered and was attributed to HGT of the entire operon, while approximately 20 operons were probably formed. The method used to identify deletion of intervening genes overlooks few cases; such genes are not identified only if all the orthologs of the intervening genes had been lost in all the species. By contrast the MinGeneCIDE method selects HT genes conservatively due to the rejection of all ambiguous cases and may well have left many HT genes unidentified. Therefore, it is likely that HGT accounts for most operon creation events. As some genes were found to be inserted in existing operons, a gene may also be inserted just downstream of another gene, forming an operon *de novo* (Fig. 6*d*). Possibly many operons are created by this mechanism early in bacterial evolution and are maintained, giving rise to conserved operons currently observed. The conclusion that most operons are created by HGT is at variance with a recent report (8). The disagreement is partly attributable to their use of predicted operons in contrast to our exclusive usage of experimentally confirmed ones. A more crucial difference is that the MinGeneCIDE method based on analyses of gene cluster alterations identifies HT genes without significant errors, while their HT identification method hinges on analyses of individual genes and provides no reliability test. We therefore consider our conclusion more probable.

**Conclusion**

   Although the co-regulation model (4) explains conserved operons in general, it is incompatible with the insertion of seemingly unrelated genes in the *relA* operon. While the apparent self-centeredness of nonconserved operons is consistent with the selfish operon model (6), the expansion of an existing operon by insertion of HT genes (Fig. 6*c*) is not. Considering the general importance of the genes and the lack of it in the conserved and nonconserved operons, respectively, we propose that genes with related essential functions tend to reside in conserved operons, whereas genes in nonconserved operons generally conferring slight advantage to the organisms undergo frequent horizontal transfer and decay.

## Materials and Methods

**Data and statistical analyses.** All the sequence data used in this study were taken from those of the three domains of life in the GTOP database (21) (October 26, 2005 version). Among the *Escherichia*, *Shigella*, and *Salmonella* species in GTOP, we analysed only the ESS species. The list of essential genes in ecol0 were obtained from http://www.shigen.nig.ac.jp/ecoli/pec/index.jsp, while functional assignments of the genes were taken from version 2.3 of the Comprehensive Microbial Resource (24), disregarding the following categories: hypothetical proteins, unclassified, and unknown function. The expectation value in each function is the product of the total number of all HT gene candidates with the functional annotation and the fraction of all genes with the functional annotation. We calculated the fraction of essential genes in each group after excluding unclassified genes. All the statistical significances were evaluated by the chi-squared test.

**Identification of orthologs.** Syntenic regions were identified by a program employing the dynamic programming algorithm based on the similarity of chromosomally encoded gene (32). We gave first preference to syntenic regions consisting of over 99 genes and second preference to syntenic regions containing between 10 and 99 genes. Syntenic regions multiply identified in the same preference category were neglected. Intergenetic regions with homology to genes in the other ESS species were searched by the Mummer program 3.0 (33) and designated as GHRs. Genes and GHRs aligned in syntenic regions were considered as orthologs.

**The MinGeneCIDE method.** We regard a continuous stretch of genes with an identical presence/absence pattern among the ESS species as a reduced ortholog cluster (abbreviated as ROC). Due to the limitation in computational power, any section with more than eight ROCs (which requires more than eight bits to describe each state) between the conserved orthologs is divided into sections of eight, except for the last one which can have less than eight, with an overlap of four from one section to the next. Although the nonconserved ROCs before the first conserved ROC and those after the last conserved ROC are not placed between conserved ROCs, we also processed them similarly. The 3,313 nonconserved ROCs were thus divided into 1,342 sections. For each ROC, all the possible states at each node are determined from those at the two subjacent node(s)/species. As the possible states at a node generally depend on those of the lower node(s), we keep the record of possible states at each node so that indels can be enumerated at the end. In the example of Fig. 1$c$, both states 11 and 01 at node G are consistent with the states of ecol0 and ecol1. If the states at nodes B and C are 11 and 00, respectively, the possible states at node A are 11, 10, 01, and 00. After determining all the possible states at node A with corresponding states at other nodes, we select the pathways that involve the minimum number of indels. For instance, the transition from 00 to 01 signifies an insertion of the first ROC, while the 10-00 transition corresponds to a deletion of the second ROC. Deletion or insertion of contiguous gene regions is regarded as one indel. For example, the 11-00 transition is considered to involve one deletion. After identification of all the possible states at node A, the number of events for each pathway is determined. If a particular bit indicates an insertion in all the possible minimum event pathways, the corresponding ROC is judged as inserted. In case the same ROC belongs to two sections as happens when the number of ROCs between conserved ROCs exceeds eight, both results must be consistent for the cluster to be regarded as inserted. The inserted genes are regarded as HT genes if a majority of the best BLAST hits in GTOP of the genes in an inserted ROC excluding those of orthologous genes are of genes in non-ESS species.

**Operon conservation.** We examined the orthologs of the genes in each operon in the Operon Data Library (34) and judged the operon conserved if and only if all the orthologs are present, are contiguous, and are in the same orientation in all of the ESS species, neglecting over-annotated genes.

We thank Dr. T. Kawabata and Mr. S. Sakamoto for constructing and constantly updating the GTOP database, Dr. T. Horiike for help in phylogeny, and Drs. K. Fukami-Kobayashi and Y.

Minezaki for constructive discussions. This work was supported in part by a grand-in-aid from the Ministry of Education, Culture, Sports, Science and Technology of Japan.

**Table 1. Species distribution of HT genes**

| Species | ecol0 | ecol1 | ecol3 | sfle0 | sfle1 | styp0 | styp2 | styp1 | Total |
|---|---|---|---|---|---|---|---|---|---|
| Positively HT genes | 22 | 53 | 18 | 3 | 0 | 0 | 0 | 20 | 116 |
| Candidates | 190 | 769 | 573 | 128 | 133 | 6 | 11 | 206 | 2016 |
| All chromsomally encoded genes | 4182 | 5360 | 5378 | 4181 | 4067 | 4394 | 4322 | 4451 | 36335 |

**Table 2. Probable origins of HT genes**

| Origin | Positively HT genes | | HT gene candidates | | Coverage* |
|---|---|---|---|---|---|
| | # | % | # | % | % |
| γ−Proteobacteria | 53 | 45.7 | 907 | 45.0 | 36.7 |
| Other bacteria | 37 | 31.9 | 624 | 31.0 | 42.1 |
| Archaea | 1 | 0.9 | 30 | 1.5 | 50.0 |
| Eukaryota | 14 | 12.1 | 225 | 11.2 | 66.7 |
| Unknown | 11 | 9.5 | 230 | 11.4 | 77.8 |
| Total | 116 | 100.0 | 2016 | 100.0 | 46.6 |

*The fraction of HT genes of the corresponding origin that were also so identified by a nucleotide composition method (10) .

**Table 3. Estimation of error in HT gene candidates selected according to phylogenetic tree A**

| BLAST best hit | Positively HT genes (assuming tree A) | | HT gene candidates (assuming tree A) | | Genes in conserved gene clusters | |
|---|---|---|---|---|---|---|
| | # | % | # | % | # | % |
| γ-Proteobacteria | 48 | 47.1 | 2419 | 51.8 | 17476 | 91.7 |
| Other bacteria | 29 | 28.4 | 1366 | 29.3 | 1144 | 6.0 |
| Archaea | 1 | 1.0 | 76 | 1.6 | 44 | 0.2 |
| Eukaryota | 14 | 13.7 | 404 | 8.7 | 204 | 1.1 |
| Unknown | 10 | 9.8 | 402 | 8.6 | 186 | 1.0 |
| Total | 102 | 100.0 | 4667 | 100.0 | 19054 | 100.0 |
| Error rate | 0% (assumed) | | *10.5% (estimated)* | | 100% (assumed) | |

**Table 4. Nonconserved operons**

| Operon | ecol1 | ecol3 | sfle0 | sfle1 | styp0 | styp2 | styp1 |
|---|---|---|---|---|---|---|---|
| *fix* | | NC | | NC | | | |
| *araBAD* | | | NC | | | | |
| *aceEF* | | NC | | | | | |
| *htrE* | | | NC | NC | | | NC |
| *fhuACDB* | | | | | NC | NC | |
| *codBA* | | | NC | NC | | | |
| *cynTSX* | | NC | NC | NC | NC | NC | NC |
| *lac* | | | NC | NC | NC | NC | NC |
| *mhp* | | NC | NC | NC | NC | NC | NC |
| *cyoABCDE* | | | NC | NC | | | |
| *copRS* | | | | | NC | NC | NC |
| *entCEBA* | | | NC | NC | | | |
| *nagBACD* | | | NC | NC | | | |
| *speFpotE* | | | NC | NC | NC | NC | |
| *kdpABC* | | | NC | NC | | | |
| *hya* | | | | | NC | NC | |
| *torCAD* | | | NC | NC | NC | | |
| *csgDEFG* | | | NC | NC | | | |
| *csgBA* | | | NC | NC | | | |
| *flgBCDEFGHI* | | | NC | NC | | | |
| *flgKL* | | | NC | NC | | | |
| *fabDGacpP* | | NC | | | | | |
| *oppABCDF* | | | | | NC | NC | |
| *trp* | | NC | | | | | |
| *sap* | | | | | NC | NC | |
| *narZYWV* | | | NC | NC | | NC | |
| *fdnGHI* | | | | | NC | NC | |
| *hip* | NC | NC | NC | NC | NC | NC | NC |
| *dicB* | | NC | NC | NC | NC | NC | NC |
| *rst* | | | NC | NC | NC | NC | NC |
| *uidABC* | | | NC | NC | NC | NC | NC |
| *malXY* | | | | | NC | NC | NC |
| *pheST* | | NC | | | | | |
| *celABCDF* | | | NC | NC | | | |
| *edd-eda* | | | | | NC | | |
| *flhBA* | | | NC | NC | | | |
| *cheRBYZ* | | NC | | | NC | NC | NC |
| *flhDC* | | | NC | NC | | | |
| *otsBA* | | | NC | NC | | | |
| *araFGH* | | | NC | NC | NC | NC | NC |
| *fliAZY* | | | NC | NC | | | |
| *fliFGHIJK* | | | NC | NC | | | |
| *fliLMNOPQR* | | | | NC | | | |
| *cobUST* | | | | NC | | | |
| *rfb* | NC | NC | NC | NC | NC | NC | NC |
| *gat* | | | NC | NC | NC | NC | NC |
| *mglBA* | | | | | NC | NC | |
| *atoSC* | NC | | NC | NC | NC | NC | NC |
| *atoDAB* | NC | | NC | NC | NC | NC | NC |
| *purF* | | | | | NC | NC | NC |
| *dsdXA* | NC | | | NC | | | |
| *xapAB* | NC | | NC | NC | | | |
| *pts* | | | | | NC | | |
| *cysPTWAM* | | | NC | | | | |
| *amiAhemF* | | | | NC | | | |
| *proVWX* | | | | | NC | NC | |
| *srlABD* | NC | | NC | NC | NC | NC | NC |
| *ascFB* | | NC | NC | NC | NC | NC | NC |
| *hyc* | | | NC | NC | | | |
| *relA* | | NC | NC | NC | NC | NC | NC |
| *fucPIK* | | NC | NC | NC | | | NC |
| *prfBlysS* | | | NC | | | | |
| *glc* | NC | NC | NC | NC | NC | NC | NC |
| *ebgAC* | | | | | NC | NC | NC |
| *uxaCA* | | | | | NC | NC | NC |
| *tdc* | | | | | NC | NC | |
| *aga* | NC | NC | NC | NC | NC | NC | NC |
| *rplUrpmA* | | NC | | | | | |
| *gltBDF* | NC | NC | NC | NC | NC | NC | NC |
| *acrEF* | | | NC | NC | NC | NC | |
| *spc* | | NC | | | | | |
| *ugpBAECQ* | | | NC | NC | | | |
| *nikABCDE* | | | | | NC | NC | NC |
| *arsBC* | | NC | NC | | NC | NC | NC |
| *xylFGH* | | | NC | NC | NC | NC | NC |
| *rfa* | NC | NC | NC | NC | NC | NC | NC |
| *glvCBG* | NC | NC | NC | NC | NC | NC | NC |
| *dnaA* | | NC | | | | | |
| *tnaAB* | | | | | NC | NC | NC |
| *bglGFB* | NC | | NC | NC | NC | NC | NC |
| *rbs* | | | NC | NC | | | |
| *frv* | | NC | | | NC | NC | NC |
| *rhaBAD* | | | | | NC | NC | |
| *thiCEFGH* | NC | | NC | NC | | | |
| *ace* | | | NC | NC | | | |
| *phn* | | NC | NC | NC | NC | NC | NC |
| *melAB* | | NC | | | | | |
| *cad* | | | NC | NC | | | |
| *rpsF-rplI* | | NC | NC | | | | |
| *chpSB* | | NC | NC | NC | NC | NC | NC |
| *treBC* | | | | | | | NC |
| *fecABCDE* | NC | NC | NC | NC | NC | NC | NC |
| # Nonconserved | 15 | 29 | 58 | 58 | 49 | 47 | 38 |

NOTE--NC signifies nonconservation of the operon in the species.

**Figure legends**

**Fig. 1.** Identification of inserted genes by the MinGeneCIDE method. (*a*) ROCs are represented by isochromatic arrows of proportional lengths presented in the ecol0 gene order with the arrowheads signifying orientations. Orthologs are vertically aligned. Lines and dotted lines respectively indicate intergenetic regions and the lack of corresponding sequences. (*b*) The same ROCs as in (*a*) in table formats with cells colored according to the ROCs. (*c*) Under each node and species in the phylogenetic tree, all the possible states are shown in bit representation. Asterisks and crosses mark the states of the two minimal event pathways.

**Fig. 2.** Alternative phylogenetic tree from ortholog sequences. The phylogeny of the ESS species based on ortholog sequences was inferred using all the genes in the conserved ROCs. The tree is topologically identical to a previously reported one (35). The *Shigella* branches are shown in green.

**Fig. 3.** Verification of gene insertion by the INTEGRAL method. ROCs are drawn as in Fig. 1*b* with the genes in starred species depicted with their orientations reversed. (*a*) The *relA* operon with the phylogenetic tree. (*b*) Conceptual scheme for verification shown with matching intergenetic regions aligned vertically. The intergenetic regions between W's and Z's are first aligned by blastn and, if aligned, the alignments are extended by clustal W. Insertion to sp. 1 is verified if sp. 3 forms an outgroup and N2 and N3 exactly line up. The intergenetic regions may have deletions.

**Fig. 4.** Functional distribution of HT genes. All the categories annotated to more than 0.5% of the ESS genes are shown. Red and green rectangles respectively indicate that the fold increases of the categories are significantly higher and lower than unity at $P < 0.01$.

**Fig. 5.** A nonconserved but possibly functional operon despite a gene loss. Genes are presented as in Fig. 3, except that GHRs are presented in a different color. All the ecol0 genes depicted reside in the *rhaBAD* operon.

**Fig. 6.** Models of operon formation and expansion. Genes are represented by arrows on circular chromosomes.

**Fig. 7.** Examples of nonconserved operons. Genes are drawn as in Fig. 3. All the ecol0 genes depicted reside in operons. (*a*) The genes in the *rst* operon and their orthologs together with the phylogenetic tree. (*b*) The genes in the *aga* operon genes and their orthologs. There are no orthologs in styp0, styp2, and styp1.

# Fig. 1

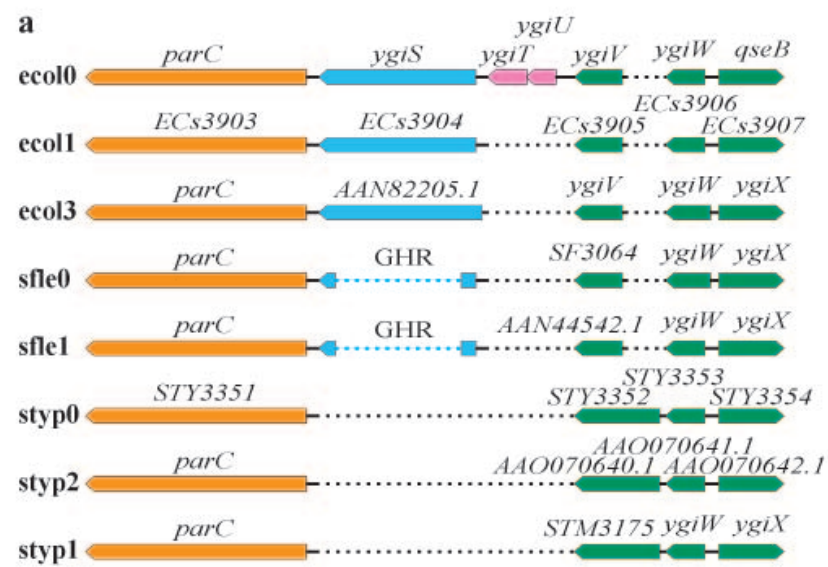

# Fig. 3

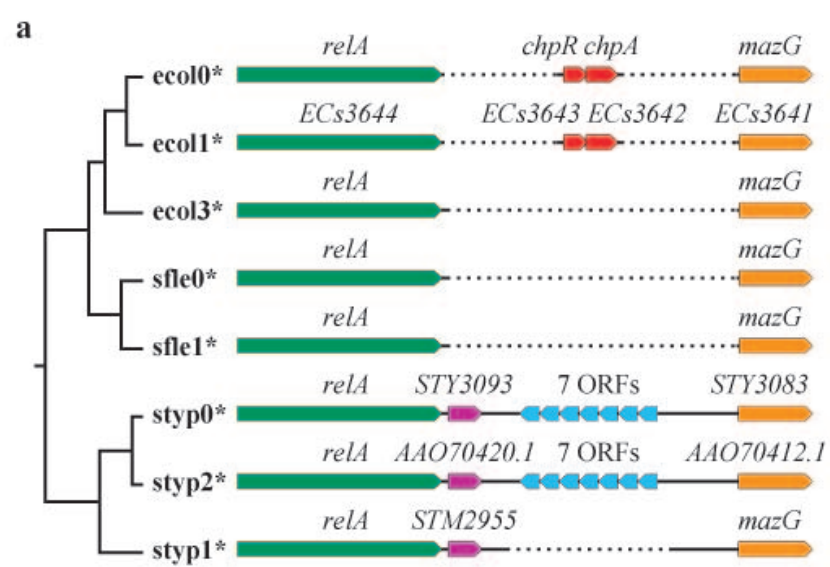

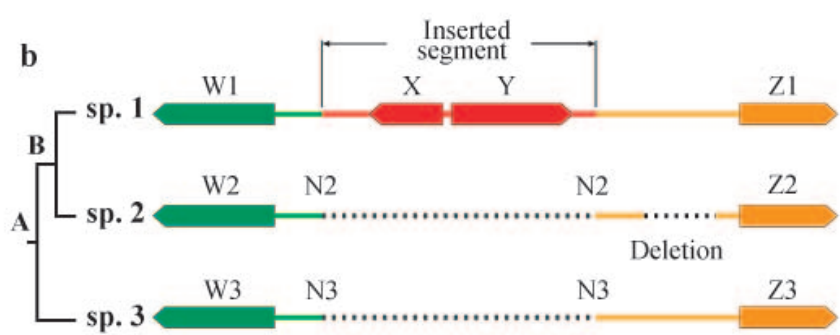

# Fig. 2

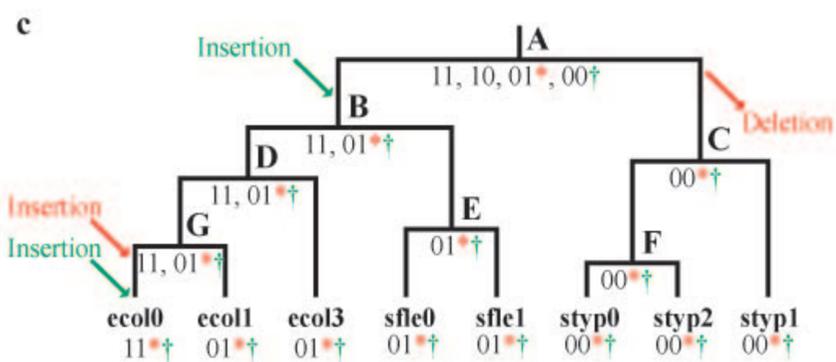

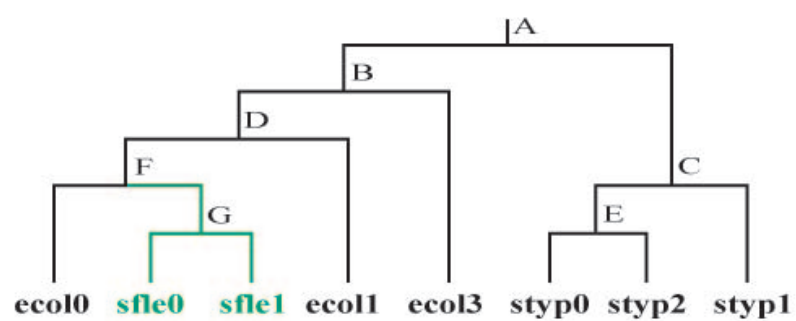

# Fig. 4

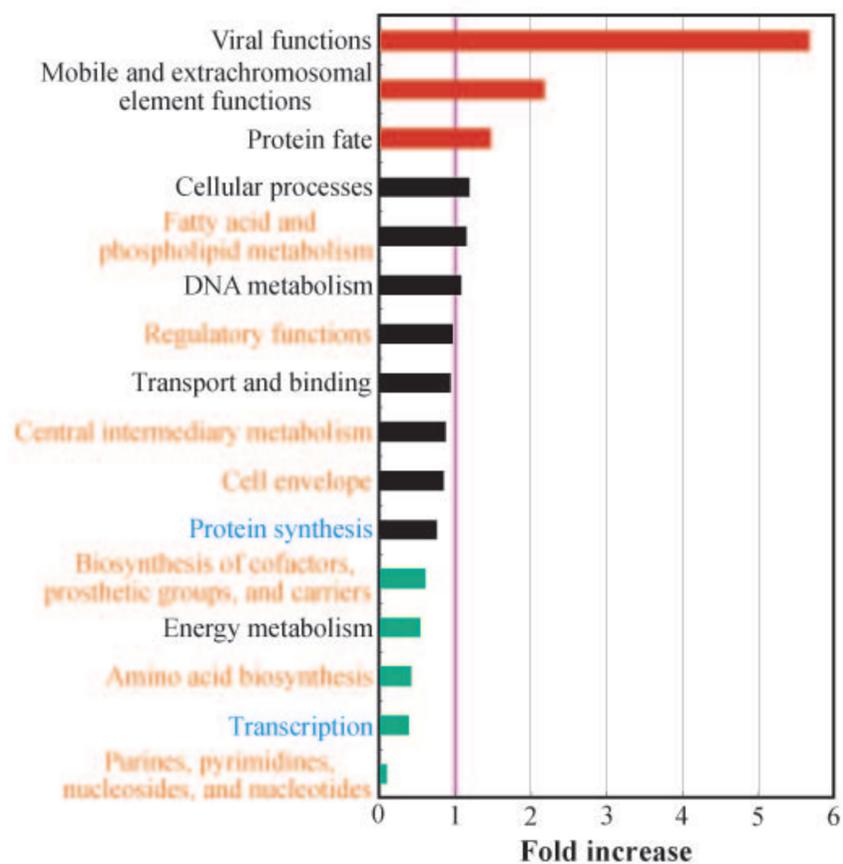

## Fig. 5

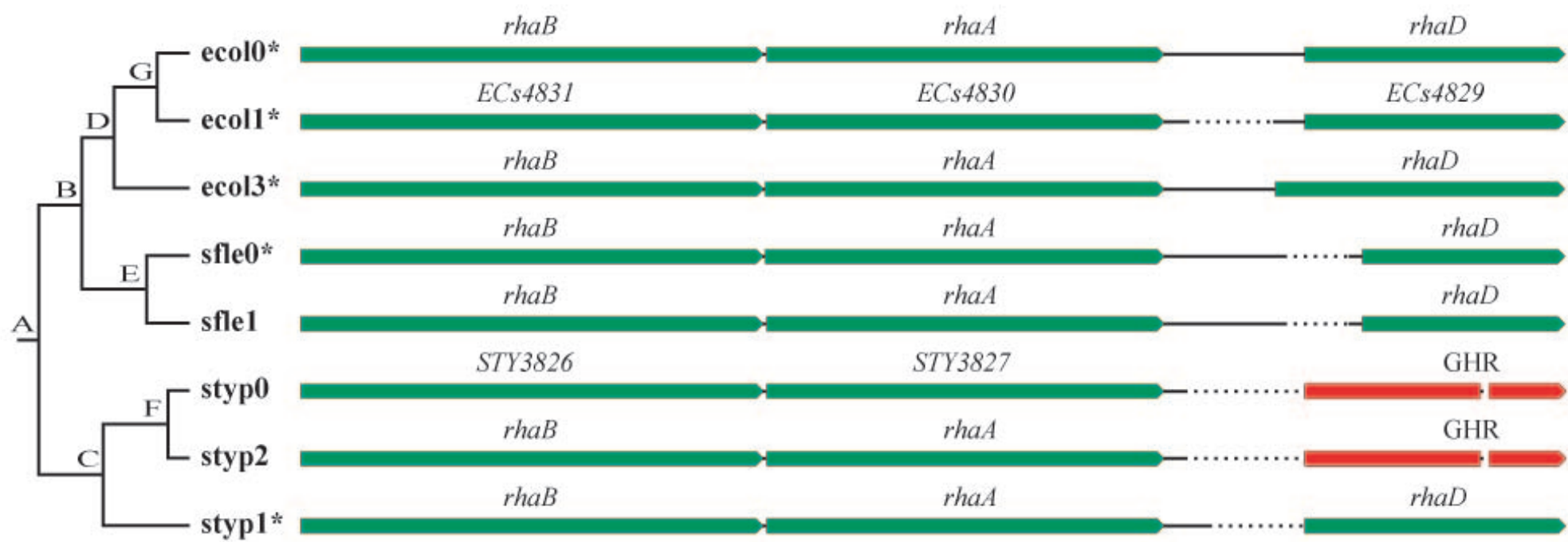

## Fig. 6

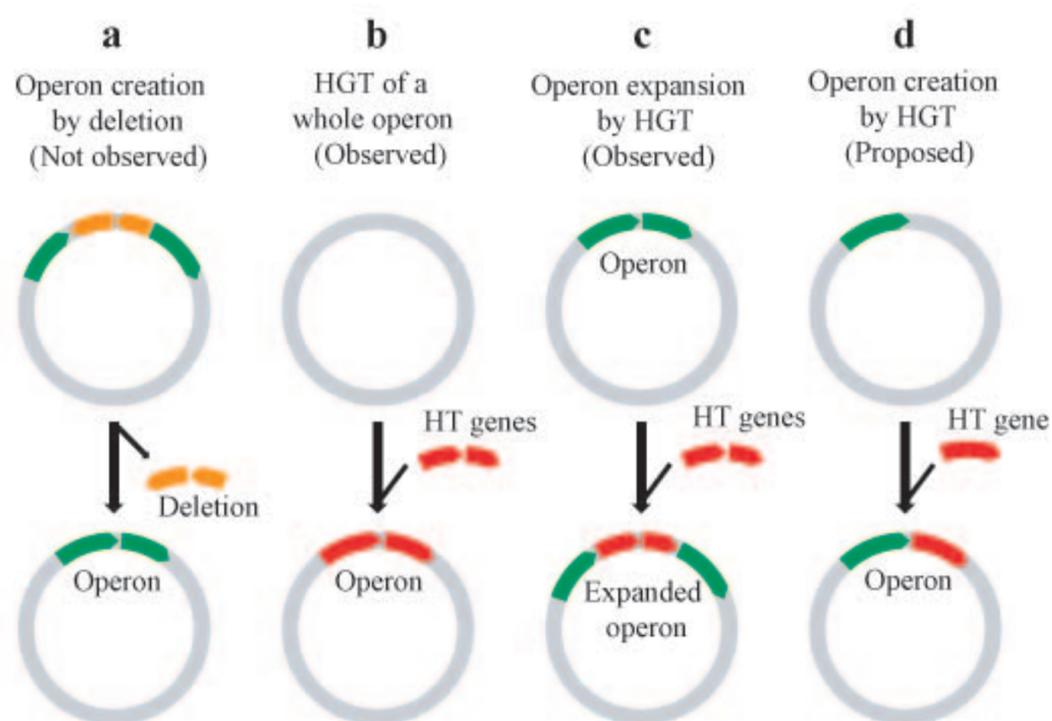

a Operon creation by deletion (Not observed)

b HGT of a whole operon (Observed)

c Operon expansion by HGT (Observed)

d Operon creation by HGT (Proposed)

## Fig. 7

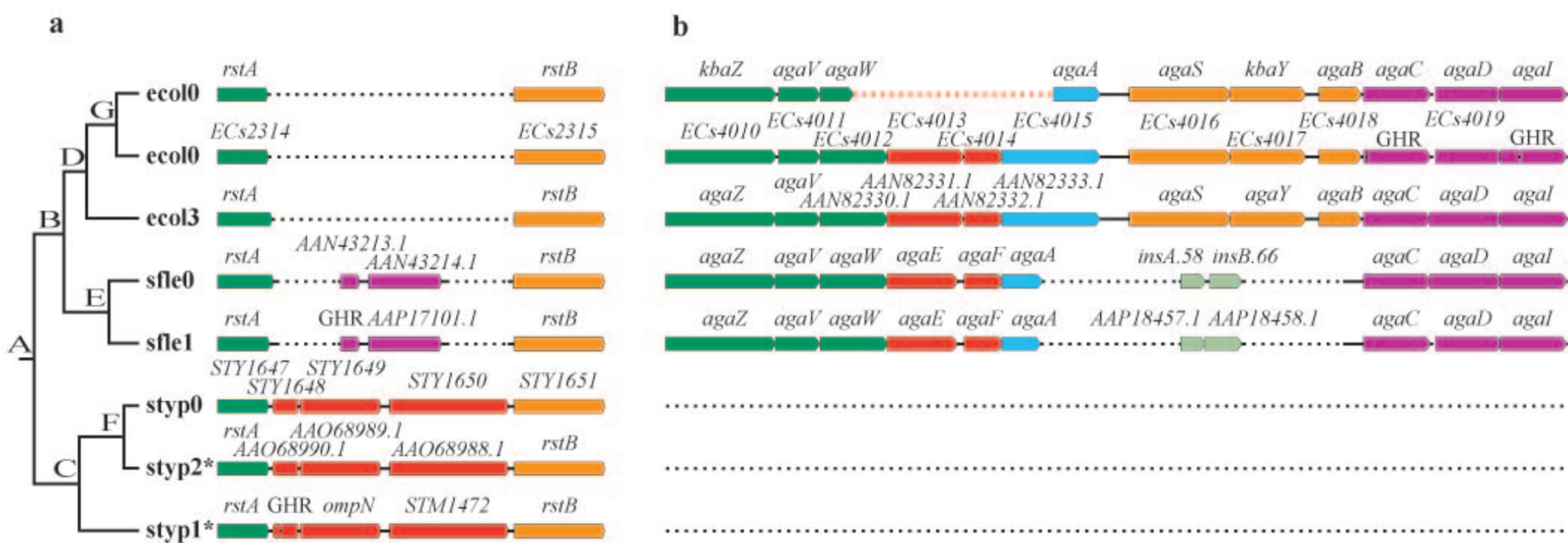